# Stability and thermoelectric properties of transition metal silicides from first principles calculations

P. Jund[*], X. Tao, R. Viennois, C. Colinet and J.-C. Tedenac

Institut Charles Gerhardt Montpellier, place Eugène Bataillon, Université Montpellier II, F-34095 Montpellier, France

**Abstract**

We report an ab-initio study of the stability and electronic properties of transition metal silicides in order to study their potential for high temperature thermoelectric applications. We focus on the family $M_5Si_3$ (M = Ta, W) which is stable up to about 2000 °C. We first investigate the structural stability of the two compounds and then determine the thermopower of the equilibrium structure using the electronic density of states and Mott's law. We find that $W_5Si_3$ has a relatively large thermopower but probably not sufficient enough for thermoelectric applications.

**Introduction**

In the search of new sustainable energies, thermogeneration of electricity is one of the promising ways, notably because it is able to produce electricity from different kinds of heat wastes and to its portability. However, despite recent progress in the field, the state-of-the art materials have specificities making their mass-production difficult or even impossible: toxicity (thallium, chalcogen or/and pnictogen based alloys), weak abundance (tellurium based alloys) or cost (germanium or rare-earth based alloys). For all these reasons, finding thermoelectric materials without these drawbacks is vital if we wish to use high temperature thermoelectricity for generating electricity in the near future. Due to their abundance, low cost and low toxicity, transition metal silicides are candidate materials of primary interest and the slightest advance in this field would immediately find an application for the thermogeneration of electricity.

* corresponding author : pjund@um2.fr

However, until now, the best transition metal silicide compound was found to be the higher manganese silicide (HMS) compound with a dimensionless figure of merit ZT = 0.5 to 0.7 at about 500 °C [1]. But the problems of stability of the HMS and the lack of knowledge of the Mn-Si phase diagram are very challenging. This shows that the stability issues are also of fundamental interest in the search of new materials for high temperature applications. It is worth noting that no study has focused on the potential for thermoelectricity of transition metal silicides with a low Si content such as $M_5Si_3$ (M=Ta,W), which we have recently begun to study [3,4]. Thus, the aim of the present study is to determine the equilibrium structure of $Ta_5Si_3$ and $W_5Si_3$ which are stable above 2000C [3,4] and to determine their thermopower using Mott's law.

**Computational details**

First-principles calculations are performed by using the scalar relativistic all-electron Blöchl's projector augmented-wave (PAW) method [5,6] within the generalized gradient approximation (GGA), as implemented in the highly-efficient Vienna ab initio simulation package (VASP) [7,8]. In the standard mode, VASP performs a fully relativistic calculation for the core-electrons and treats valence electrons in a scalar relativistic approximation. For the GGA exchange- correlation function, the Perdew-Wang parameterization (PW91) [9, 10] is employed. Here we adopted the standard version of the PAW potentials for Si, W and Ta atoms. A plane-wave energy cutoff of 450 eV is held constant for the study of the different $M_5Si_3$ compounds. Brillouin zone integrations are performed using Monkhorst-Pack k-point meshes [11], and the Methfessel-Paxton technique [12] with a smearing parameter of 0.2 eV. The reciprocal space (k-point) meshes are increased to achieve convergence to a precision of 1meV/atom. The total energy is converged numerically to less than $1\times10^{-6}$ eV/unit with respect to electronic, ionic and unit cell degrees of freedom, and the latter two are relaxed using calculated forces with a preconditioned conjugated gradient algorithm. After structural optimization, calculated forces are converged to less than 0.01eV/Å. All calculations are performed using the "high" precision setting within the VASP input file to avoid wrap-around errors.

**Results**

First, we discuss the stability of the transition metal silicides considered in this study. Three crystallographic structures have been considered: the $W_5Si_3$-prototype structure ($D8_m$), the $Cr_5B_3$-prototype structure ($D8_l$) and the $Mn_5Si_3$-prototype structure ($D8_8$). The calculated lattice constants and several elastic constants can be found in ref [3] and [4] and are not recalled here but in Fig. 1 one can see that the equilibrium structure of $Ta_5Si_3$ is the $D8_l$ structure while the $D8_m$ structure is the equilibrium structure of $W_5Si_3$ both having the body centered tetragonal structure with space group I4/mcm. These results are in agreement with the available experimental data as well as the other simulated structural data [3] and [4].

Once the equilibrium structures have been determined, the next step is to calculate the partial and total electronic density of states (TDOS). The latter is shown in Fig. 2 in the energy region close to the Fermi level for both compounds (the detailed DOS can be found in ref [3] and [4]). Our aim here is to obtain the thermopower α of $Ta_5Si_3$ and $W_5Si_3$ from the TDOS, using the well known Mott's law [13,14]:

$$\frac{\alpha}{T} = -\frac{\pi^2 k_B^2}{3e}\left(\frac{\partial \ln(\sigma(\varepsilon))}{\partial \varepsilon}\right)_{\varepsilon_F} \quad (1)$$

where T is the temperature and σ is the electrical conductivity.

If we assume a negligible k-dependence of the group velocity close to the Fermi level and with the hypothesis of constant relaxation time, we get:

$$\frac{\alpha}{T} = -\frac{\pi^2 k_B^2}{3e}\left(\frac{\partial \ln(n(\varepsilon))}{\partial \varepsilon}\right)_{\varepsilon_F} \quad (2)$$

where n(ε) is the TDOS.

As discussed by Tobola and coworkers [14], in favorable cases, eq. (2) can give qualitative information on the doping dependence of the thermopower and permits, prior to any experimental measurement, to predict the thermopower of different compounds as is the case here.

From Figure 2, we see that the TDOS close to the Fermi level is varying faster for the tungsten compound than for the tantalum one. This indicates that $W_5Si_3$ must be better than $Ta_5Si_3$ for thermoelectric applications. Indeed, from Mott's relation (eq. 2), we find a very low thermopower for $Ta_5Si_3$ (-1.8 µV/K at 300 K) similar to normal metals whereas $W_5Si_3$ has a much larger thermopower (-17 µV/K at 300 K). Although these informations are mainly qualitative, they indicate that only $W_5Si_3$ could be interesting for its thermoelectric properties at high temperature, although the thermopower seems still too low. This thermopower can probably be increased with doping.

**Conclusion**

From ab-initio calculations, we have studied the potential of some transition metal silicides for high temperature thermoelectric applications. First we have determined the stable structure of $Ta_5Si_3$ and $W_5Si_3$ using accurate total energy calculations. In agreement with experimental data we have found the $D8_l$ structure as being the equilibrium structure of $Ta_5Si_3$ while $W_5Si_3$ crystallizes in the $D8_m$ structure. Using Mott's law and after having calculated the total density of states of the two compounds, we have determined their thermopower at 300K. We have found that $Ta_5Si_3$ is a poor thermoelectric compound whereas $W_5Si_3$ has a thermopower one order of magnitude larger than the one found for $Ta_5Si_3$. Nevertheless this thermopower is still relatively small and further investigations are necessary to explore the effect of doping.

**Figures**

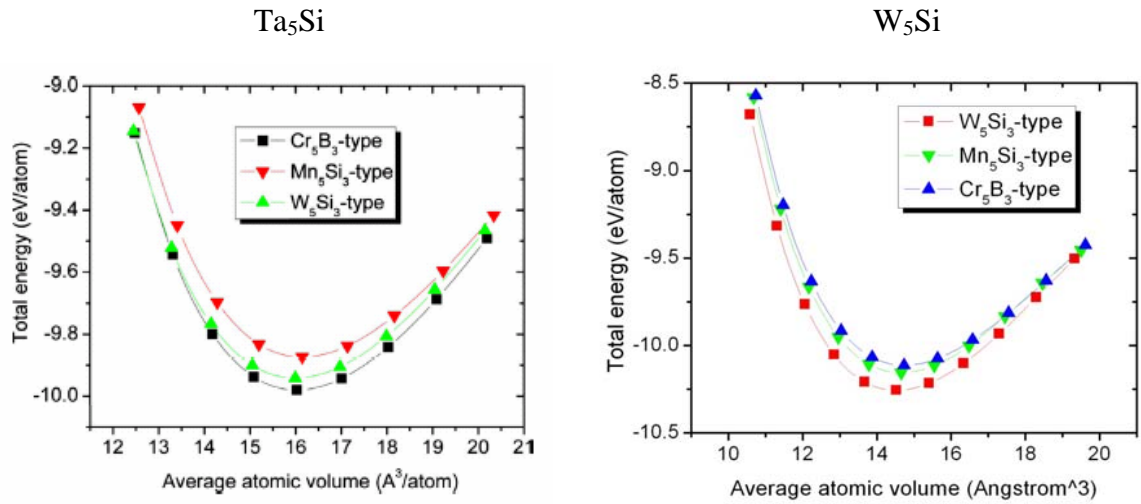

Figure 1: Total energy vs volume for the $Cr_5B_3$, $W_5Si_3$ and $Mn_5Si_3$-prototype structures of $Ta_5Si_3$ and $W_5Si_3$

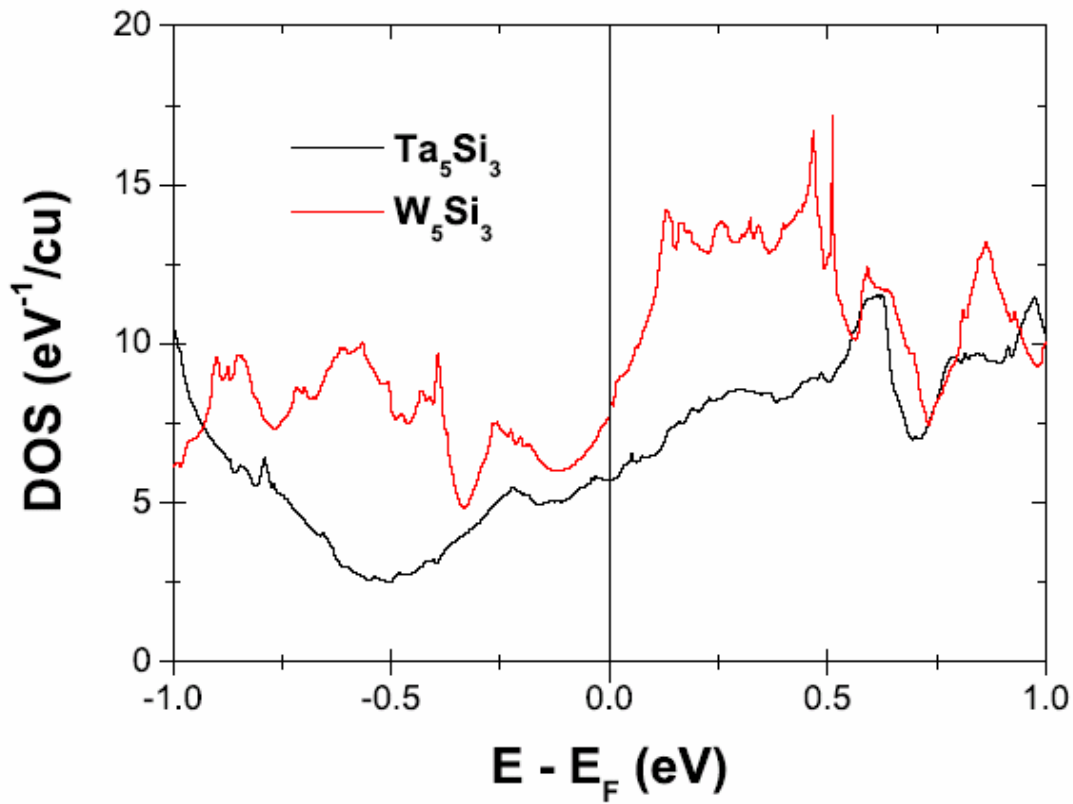

Figure 2: Total electronic density of states of $Ta_5Si_3$ and $W_5Si_3$ close to the Fermi level.